\documentclass[spmi]{academic}
\usepackage{epsfig,amsmath}

\begin{document}

\shortauthor{Mortensen {\it et al.}}

\shorttitle{Dephasing in semiconductor--superconductor ...}

\title{ Dephasing in semiconductor--superconductor structures by
coupling to a voltage probe }

\author[1]{Niels Asger Mortensen}

\author[1]{Antti-Pekka Jauho}

\author[2]{Karsten Flensberg}

\address[1]{Mikroelektronik Centret, Technical University of
  Denmark, \O rsteds Plads, Bld. 345 east, DK-2800 Kgs. Lyngby, Denmark}

\address[2]{\O rsted Laboratory, Niels Bohr
Institute, University of Copenhagen, Universitetsparken 5, DK-2100 Copenhagen \O, Denmark}


\maketitle

\begin{abstract}
We study dephasing in semiconductor--superconductor structures
caused by coupling to a voltage probe. We consider structures
where the semiconductor consists of two scattering regions between
which partial dephasing is possible. As a particular example we
consider a situation with a double-barrier junction in the normal
region. For a single-mode system we study the conductance both as
a function of the position of the Fermi level and as a function of
the barrier transparency. At resonance, where the double-barrier
is fully transparent, we study the suppression of the ideal
factor-of-two enhancement of the conductance when a finite
coupling to the voltage probe is taken into account.
\end{abstract}

\section{Introduction}

Charge transport through a normal conductor--superconductor (NS)
interface is accompanied by a conversion of quasiparticle current
to a supercurrent. In the Andreev reflection, by which the
conversion occurs, an electron-like quasiparticle in the normal
conductor (with an excitation energy lower than the energy gap of
the superconductor) incident on the NS interface is retroreflected
into a hole-like quasiparticle (with reversal of its momentum and
its energy relative to the Fermi level) and a Cooper pair is added
to the condensate of the superconductor \cite{andreev64}. For an
ideal NS interface, a signature of Cooper pair transport and the
Andreev scattering is a doubling of the conductance compared to
the normal state conductance.

A theoretical framework for studying scattering at NS interfaces
is provided by the Bogoliubov--de Gennes (BdG) formalism
\cite{degennes66} where the scattering states are eigenfunctions
of the BdG equation which is a Schr\"{o}dinger-like equation in
electron-hole space (Nambu space).

For a phase-coherent normal region, the conducting properties can
be found using a scattering matrix approach based on the BdG
equation. In the linear-response regime in zero magnetic field,
Beenakker \cite{beenakker92} found that the sub-gap conductance
$G\equiv\partial I/\partial V$ is given by
\begin{equation}
G_{\scriptscriptstyle\rm NS} = 2 G_0 {\rm Tr}\, \left(t t^\dagger
\left[\hat{2} -tt^\dagger\right]^{-1} \right)^2 =2G_0
\sum_{n=1}^{N} \frac{T_n^2}{\left(2-T_n\right)^2},
\label{BEENAKKER}
\end{equation}
which, in contrast to the Landauer formula \cite{landauer57,landauer70,fisher81,buttiker86a} for the
normal state conductance,
\begin{equation}
G_{\scriptscriptstyle\rm N} = G_0 {\rm Tr}\, t t^\dagger =G_0
\sum_{n=1}^{N} T_n, \label{LANDAUER}
\end{equation}
is a non-linear function of the transmission eigenvalues $T_n$
($n=1,2,\ldots,N$) of $t t^\dagger$. Here $G_0=2e^2/h$ is the
quantum unit of conductance for a spin-degenerate system and $t$
is the $N\times N$ transmission matrix of the normal region, $N$
being the number of propagating modes. The conductance formula
holds for an arbitrary disorder potential and is a multi-channel
generalization of a conductance formula first obtained by Blonder,
Tinkham and Klapwijk \cite{blonder82} who considered a delta
function potential as a model for the interface barrier potential.
Eq. (\ref{BEENAKKER}) is computationally convenient since standard
methods developed for quantum transport in normal conducting
semiconductor structures can be applied to calculate the
transmission matrix $t$ or the corresponding transmission
eigenvalues $T_n$.

Equations (\ref{BEENAKKER}) and (\ref{LANDAUER}) apply to
two-probe structures where the length $L$ of the disordered normal
region is much shorter than the phase-correlation length over
which quasiparticle propagation remains phase-coherent.

\begin{figure}
\begin{center}
\epsfig{file=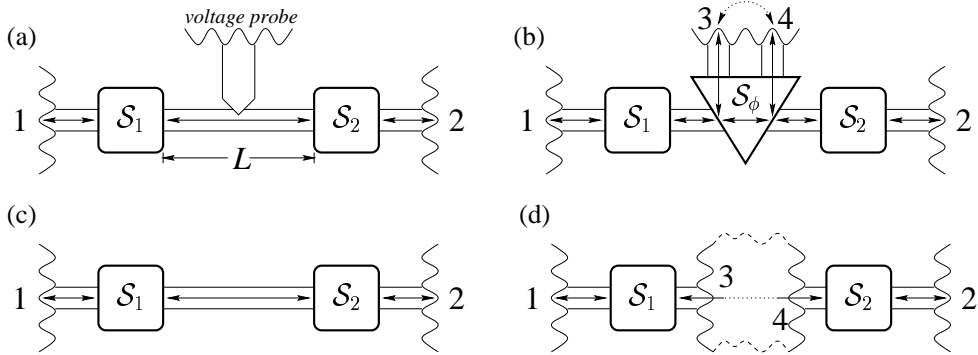, width=0.9\columnwidth}
\end{center}
\caption[]{Scattering scheme for a two probe structure with two
  scattering regions (characterized by the S--matrices ${\cal S}_1$
  and ${\cal S}_2$) between which dephasing is caused by coupling to a voltage probe. The left probe (1) is in the normal state and the
  right probe (2) is a metal which can be either in the normal or
  superconducting state. The voltage probe could {\it e.g.} be an STM tip. Panel (a) shows the generic structure, (b) corresponds to the present model for the voltage probe, (c) to the phase-coherent
  regime with no coupling to the voltage probe, and (d) to the sequential tunneling regime with full coupling to the voltage probe. The arrows
  indicate possible directions of current flow and the arrows with
  dotted midsections indicate equilibration of quasiparticles.}
\label{FIG1}
\end{figure}

The effect of coupling a mesoscopic region to a voltage probe was
first used by B\"{u}ttiker in the context of the role of quantum
phase-coherence in normal conducting series resistors
\cite{buttiker86b} and subsequently in the study of the cross-over
from coherent to sequential tunneling in series of barriers
\cite{buttiker88}. In the work of Chang and Bagwell \cite{chang97} the model of B\"{u}ttiker \cite{buttiker88} was used as a normal-metal probe controlling the Andreev-level occupation in a Josephson junction. In this paper we apply the model of B\"{u}ttiker to a two-probe semiconductor-superconductor junction with a phase-coherent normal region. However, here we consider the case
where the dephasing reservoir is constituted by a real voltage
probe. As a general model we consider a normal region with two
arbitrary scattering regions (characterized by scattering matrices
$S_1$ and $S_2$) separated by a coherent ballistic conductor of
length $L$ to which the voltage probe couples, see (a) in Fig.
\ref{FIG1}.

Specifically we study the case where the total current is
conserved, but, due to the coupling to the voltage probe the
non-equilibrium distribution of electrons and holes equilibrates
to a Fermi--Dirac distribution. In Nambu space this corresponds to
a decay of electron-hole states in the normal region. As a
particular example we consider a semiconductor-superconductor
structure with a normal region containing a double-barrier
junction of length $L$ and model the cross-over from vanishing
dephasing to full dephasing in the conducting properties.

We note that in our approach the voltage probe is real as it is
also the case in the recent work of Gramespacher and B\"{u}ttiker
\cite{gramespacher99} where a tunneling tip is coupled weakly to
the normal side of a two-probe NS structure. As in the original
work of B\"{u}ttiker \cite{buttiker86b,buttiker88} the voltage
probe could also be considered as a fictitious probe used in
modeling a finite dephasing length of the normal region. However,
we note that the model only accounts for dephasing caused by
inelastic scattering and that thermal dephasing, which is the
dominant dephasing mechanism in most semiconductor--superconductor
structures, can not be accounted for by this model
\cite{beenakker99}. Dephasing due to a finite temperature can in principal be incorporated by calculating the full energy dependence of the S--matrices, see {\it e.g.} \cite{lambert93,beenakker95,beenakker97,lambert98}, or through Green functions methods, see {\it e.g.} \cite{lambert98} and references therein.  
   
The paper is organized as follows: In Section II the S--matrix
formalism is introduced, in Section III we formulate our general
model and present the corresponding scattering scheme. In Section
IV we present results of an application of our scattering scheme
to the problem of double-barrier tunneling. Finally, in Section V
discussion and conclusions are given.

\section{Scattering matrix formalism}

Using the B\"{u}ttiker voltage probe \cite{buttiker88} we need to
consider four-probe structures (two real and two probes accounting
for the voltage probe) and we therefore need multi-probe
generalizations of Eqs. (\ref{BEENAKKER}) and (\ref{LANDAUER}).
The scattering approach to dc transport in superconducting hybrids
follows closely the scattering theory developed for
non-superconducting mesoscopic structures, see {\it e.g.} the monograph
of Datta \cite{datta95}. For a multi-probe system, the normal
state current is given by the multi-probe conductance formula of
B\"{u}ttiker \cite{buttiker86a} which in the linear-response limit
gives the following current in lead $p$:

\begin{equation}\label{buttikerN}
{\cal I}_p=\sum_{q=1}^M {\cal G}_{pq}\left[ V_p-V_q \right]\,,\,
{\cal G}_{pq}=G_0 {\rm Tr}\, {\cal
  S}_{pq}{\cal S}_{pq}^\dagger,
\end{equation}
where $q=1,2,3,\ldots, M$ labels the $M$ probes and $V_q$ is the
potential of probe $q$. The S--matrix ${\cal S}$ is an $M\times M$
block-matrix with matrices ${\cal S}_{pq}$ describing scattering of
an electron state in lead $q$ into an electron state in lead $p$.
These matrices are $N_p\times N_q$ matrices where $N_p$ and
$N_q$ are the number of modes in lead $p$ and $q$, respectively.
In the linear-response limit, the S--matrices are evaluated at the
Fermi level.

For an NS interface with multiple normal probes,
Lambert, Hui and Robinson \cite{lambert93} considered the sub-gap
current which in the linear-response regime is given by the
formula

\begin{equation}\label{buttikerNSgeneral}
I_{p}= \sum_{q=1}^M G_{pq} [V_q-V]\,,\, G_{pq} = G_0{\rm
Tr}\,\Big(\hat{1} \delta_{pq}+ S_{pq}^{\scriptscriptstyle\rm eh}
\left\{S_{pq}^{\scriptscriptstyle\rm eh}\right\}^\dagger
-S_{pq}^{\scriptscriptstyle\rm ee}
\left\{S_{pq}^{\scriptscriptstyle\rm ee}\right\}^\dagger\Big),
\end{equation}
where $\hat{1}$ is an $N_p\times N_p$ identity matrix,
$\delta_{pq}$ is the Kronecker delta symbol and $V$ is the
potential of the superconductor. The S--matrix in electron-hole
space is given by

\begin{equation}\label{compositeS}
S=\left(\begin{array}{cc}S^{\scriptscriptstyle\rm ee} &
S^{\scriptscriptstyle\rm eh}\\ S^{\scriptscriptstyle\rm
he}&S^{\scriptscriptstyle\rm hh}
\end{array}\right),
\end{equation}
where the sub-matrix $S_{pq}^{\scriptscriptstyle\rm ee}$
($S_{pq}^{\scriptscriptstyle\rm hh}$) is for normal scattering of
an electron (hole) state in lead $q$ to an electron (hole) state
in lead $p$, and $S_{pq}^{\scriptscriptstyle\rm he}$
($S_{pq}^{\scriptscriptstyle\rm eh}$) is for Andreev scattering of
an electron (hole) state in lead $q$ to a hole (electron) state in
lead $p$.

The two-probe case ($M=2$) was studied by Takane and Ebisawa
\cite{takane92} and subsequently by Beenakker \cite{beenakker92}
who derived Eq. (\ref{BEENAKKER}) within the Andreev approximation
\cite{andreev64} and the rigid boundary condition for the pairing
potential of the superconductor. Within these approximations, the
interface acts as a phase-conjugating mirror and the interface
scattering can be described by an S--matrix which at the Fermi
level is given by \cite{andreev64,beenakker92}

\begin{equation}
\label{Andreev-matrix} S_{\rm A} = \left(\begin{array}{cc}\hat{0}
&
  e^{i\left(\varphi-\frac{\pi}{2}\right)}\hat{1}\\e^{-i\left(\varphi +
      \frac{\pi}{2}\right)}\hat{1}&\hat{0} \end{array}\right),
\end{equation}
where $\varphi$ is the phase of the pair potential of the
superconductor.

The form of the BdG equation \cite{degennes66} in the normal
region where the pairing potential $\Delta$ is zero means that the
elastic scattering in the normal region due to an arbitrary
disorder potential is characterized by the block-diagonal
S--matrix

\begin{equation}
\label{disorder-matrix} S_{\rm N} = \left(\begin{array}{cc}{\cal
S} &\hat{0} \\\hat{0} &
  {\cal S}^*\end{array}\right),
\end{equation}
where $\cal S$ is the normal state S--matrix for electrons (the
S--matrix entering {\it e.g.} Eq. (\ref{buttikerN})\,) and ${\cal S}^*$
is the corresponding S--matrix for scattering of holes. The
S--matrix in Eq. (\ref{compositeS}) entering Eq.
(\ref{buttikerNSgeneral}) is given by the composite result
$S=S_{\rm A}\otimes S_{\rm N}$ where we use the notation of Datta
\cite{datta95} with the meaning of $\otimes$ found by elimination
of internal current amplitudes.

\section{Scattering scheme}

We consider an NS structure with a normal region containing two
scattering regions

\begin{equation}\label{S_1}
{\cal S}_1
=\left(\begin{array}{cc}r_1&t_1'\\t_1&r_1'\end{array}\right)\,,\,
{\cal S}_2
=\left(\begin{array}{cc}r_2&t_2'\\t_2&r_2'\end{array}\right),
\end{equation}
connected by a ballistic conductor of length $L$. The free
propagation of electron-like quasiparticles is described by
\begin{equation}
U=\left(\begin{array}{cc}\hat{0} & X\\ X
&\hat{0}\end{array}\right),
\end{equation}
with elements given by $X_{nn'}=\delta_{nn'}\exp\left(ik_n
L\right)$. For a wire of width $W$ with a hard-wall confining
potential the longitudinal wave vector $k_n$ at the Fermi level is
given by $k_n=k_{\rm F}\sqrt{1-\left(n\pi/k_{\rm F}W\right)^2}$.

We connect voltage probe to the center of the conductor, see panel
(b) in Fig. \ref{FIG1}. Using the model of B\"{u}ttiker
\cite{buttiker88} the conductor is now characterized by the
unitary matrix

\begin{equation}
{\cal S}_{\rm \phi}= \left(\begin{array}{cccc} \hat{0} &
\sqrt{1-\zeta} X &\sqrt{\zeta}X^{1/2}&\hat{0}\\ \sqrt{1-\zeta} X &
\hat{0} &\hat{0} &\sqrt{\zeta}X^{1/2}\\ \sqrt{\zeta}X^{1/2} &
\hat{0} &\hat{0}&- \sqrt{1-\zeta}\hat{1}\\ \hat{0} &
\sqrt{\zeta}X^{1/2} &- \sqrt{1-\zeta}\hat{1}&\hat{0}
\end{array}\right),
\end{equation}
which is a generalization of the S--matrix in Ref.
\cite{buttiker88} with the matrix $X$ taking the finite length of
the conductor into account. Here $0\leq \zeta\leq 1$ is a
parameter determining the coupling of the conductor to the voltage
probe . The scattering from lead $q$ to lead $p$ is given by the
sub-matrices $\left\{{\cal
    S}_\phi\right\}_{pq}$ which are zero along the diagonal since
there is no back-scattering from the conductor and the other zeros
reflect that lead 1 couples only to leads 2 and 3, lead 2 only to
leads 1 and 4 etc. The basic idea is that current can flow either
directly through the phase-coherent conductor, as shown in panel
(c) of Fig. \ref{FIG1}, or in the sequential way, as in panel (d).
Using the scheme in panel (b), the cross-over from coherent to
sequential tunneling can be modeled by changing the coupling from
$\zeta=0$ to $\zeta=1$. In this approach, probes 3 and 4 have a
common reservoir (that of the voltage probe) and quasiparticles entering probe 3 are
re-injected through probe 4, and vice versa, so that the common
reservoir of probes 3 and 4 does not supply or draw any net
current ($I_3+I_4=0$).

The composite S--matrix of the normal region ${\cal S}={\cal
  S}_2\otimes {\cal S}_{\rm \phi}\otimes {\cal S}_1$ can now be found.
Eliminating the internal current amplitudes and introducing the
matrices $\gamma_{12}\equiv \left[\hat{1}-\left(1-\zeta\right)Xr_2
X
  r_1'\right]^{-1}$ and $\gamma_{21}\equiv
\left[\hat{1}-\left(1-\zeta\right)Xr_1'Xr_2\right]^{-1}$ we get

\begin{multline}\label{S_N}
{\cal S} = \left(\begin{array}{cc} r_1
+\left(1-\zeta\right)t_1'\gamma_{12}X r_2X t_1
&\sqrt{1-\zeta}t_1'\gamma_{12} Xt_2' \\
\sqrt{1-\zeta}t_2\gamma_{21} Xt_1
&r_2'+\left(1-\zeta\right)t_2\gamma_{21}X r_1'Xt_2'\\
\sqrt{\zeta}X^{1/2}\gamma_{12}^T t_1 &
\sqrt{\zeta\left(1-\zeta\right)}X^{1/2}r_1'\gamma_{12} X t_2'\\
\sqrt{\zeta\left(1-\zeta\right)}X^{1/2}r_2\,\gamma_{21}Xt_1
&\sqrt{\zeta}X^{1/2}\gamma_{21}^T t_2'
\end{array}\right.\\ 
\left.\begin{array}{cc} \sqrt{\zeta}t_1'\gamma_{12}X^{1/2} &
\sqrt{\zeta\left(1-\zeta\right)}t_1'\gamma_{12}X r_2 X^{1/2}\\
\sqrt{\zeta\left(1-\zeta\right)} t_2\gamma_{21}X r_1'X^{1/2}
&\sqrt{\zeta}t_2\gamma_{21}X^{1/2}\\ \zeta X^{1/2} r_1'\gamma_{12}
X^{1/2} &-\sqrt{1-\zeta}\left\{\hat{1}-\zeta X^{1/2} r_1'
\gamma_{12}
  Xr_2 X^{1/2}\right\} \\
-\sqrt{1-\zeta}\left\{\hat{1}-\zeta X^{1/2}r_2 \gamma_{21}
  Xr_1' X^{1/2}\right\} & \zeta X^{1/2} r_2 \gamma_{21}X^{1/2}
\end{array}\right),
\end{multline}
which together with Eq. (\ref{buttikerN}) gives the normal state
conducting properties.

In order to apply Eq. (\ref{buttikerNSgeneral}) for the conducting
properties of the superconducting state we need to calculate the
two block-matrices $S^{\scriptscriptstyle\rm ee}$ and
$S^{\scriptscriptstyle\rm he}$ of the composite S--matrix
$S=S_{\rm
  A}\otimes S_{\rm N}$. Since the phase $\varphi$ of the pairing
potential for a 2DEG-S system (in contrast to S-2DEG-S systems)
can be arbitrary and the conducting properties are independent of
$\varphi$, we may for simplicity let $\varphi=0$. Eliminating
internal current amplitudes we get

\begin{eqnarray}\label{See}
S^{\scriptscriptstyle\rm ee}&=&\left(\begin{array}{cccc} {\cal
S}_{11} - {\cal S}_{12} {\cal S}_{22}^* \Gamma_{22} {\cal S}_{21}
& \hat{0} &{\cal S}_{13} - {\cal S}_{12} {\cal S}_{22}^*
\Gamma_{22} {\cal S}_{23} &{\cal S}_{14} - {\cal S}_{12} {\cal
S}_{22}^*\Gamma_{22} {\cal S}_{24}\\ \hat{0} & \hat{0} &\hat{0} &
\hat{0}\\ {\cal S}_{31} - {\cal S}_{32} {\cal S}_{22}^*
\Gamma_{22} {\cal S}_{21} & \hat{0} &{\cal S}_{33} - {\cal S}_{32}
{\cal S}_{22}^*\Gamma_{22} {\cal S}_{23} &{\cal S}_{34} - {\cal
S}_{32} {\cal S}_{22}^* \Gamma_{22} {\cal S}_{24}\\ {\cal S}_{41}
- {\cal S}_{42} {\cal S}_{22}^* \Gamma_{22}{\cal S}_{21} & \hat{0}
&{\cal S}_{43} - {\cal S}_{42} {\cal S}_{22}^* \Gamma_{22} {\cal
S}_{23} &{\cal S}_{44} - {\cal S}_{42} {\cal S}_{22}^* \Gamma_{22}
{\cal S}_{24}
\end{array}\right),\\
\label{She} S^{\scriptscriptstyle\rm
he}&=&\left(\begin{array}{cccc} -i {\cal S}_{12}^* \Gamma_{22}
{\cal S}_{21}& \hat{0} & -i{\cal S}_{12}^* \Gamma_{22} {\cal
S}_{23} & -i {\cal S}_{12}^*\Gamma_{22} {\cal S}_{24}\\ \hat{0}
&\hat{0}  &\hat{0}&\hat{0}\\ -i {\cal S}_{32}^*\Gamma_{22} {\cal
S}_{21}& \hat{0} &-i {\cal S}_{32}^*\Gamma_{22} {\cal S}_{23} &-i
{\cal S}_{32}^* \Gamma_{22} {\cal S}_{24}\\ -i {\cal
S}_{42}^*\Gamma_{22} {\cal S}_{21} & \hat{0} &-i {\cal
S}_{42}^*\Gamma_{22} {\cal S}_{23} &-i {\cal S}_{42}^* \Gamma_{22}
{\cal S}_{24}
\end{array} \right),
\label{Shh}
\end{eqnarray}
where the matrix $\Gamma_{22} \equiv \left[\hat{1} + {\cal
    S}_{22}{\cal S}_{22}^*\right]^{-1}$ has been introduced. The zero
matrices in row $2$ and column $2$ reflect that the superconductor
(probe $2$) does not carry any quasiparticle current due to the
gap in the density of states. At the interface, the quasiparticle
current is transformed to a supercurrent carried by the Cooper
pairs of the condensate so that current is conserved. For a finite
bias exceeding the energy gap of the superconductor, the current
in the superconducting probe will be carried by both
quasiparticles and Cooper pairs.

To find the normal state conductance $G_{\rm N}$ we apply Eq.
(\ref{buttikerN}). From Kirchhoff's law $\sum {\cal I}_p=0$ and
the condition ${\cal I}_3+{\cal I}_4=0$ for the voltage probe get
\begin{equation}
\left(\begin{array}{c}{\cal I}_1\\ 0 \end{array}\right)= \left(
\begin{array}{cc}{\cal G}_{12} +{\cal G}_{13}+ {\cal
  G}_{14}&-{\cal G}_{13} - {\cal G}_{14}\\
-{\cal G}_{31}-{\cal G}_{41}&{\cal G}_{31} +{\cal G}_{32}+{\cal
G}_{41} +{\cal G}_{42}
\end{array}\right) \left(\begin{array}{c}V_1-V_2\\\bar{V}-V_2\end{array}\right),
\end{equation}
where $\bar{V}=V_3=V_4$ is the common potential of the voltage
probe reservoir. The conductance $G_{\rm N}=\partial {\cal
  I}_1/\partial(V_1-V_2)$ is therefore given by
\begin{equation}\label{G_N}
G_{\rm N}= {\cal G}_{12} +{\cal G}_{13}+ {\cal G}_{14} -
\frac{\left({\cal G}_{13} +{\cal G}_{14}\right) \left({\cal
G}_{31} +{\cal G}_{41}\right)} {{\cal G}_{31} +{\cal G}_{32}+{\cal
G}_{41} +{\cal G}_{42}}.
\end{equation}

In the superconducting state we apply Eq.
(\ref{buttikerNSgeneral}) and in the same way we find that
\begin{equation}
\left(\begin{array}{c}I_1\\0\end{array}\right)= \left(
\begin{array}{cc}G_{11} &G_{13}+ G_{14}\\ G_{31}+G_{41}&
G_{33}+G_{34}+G_{43}+G_{44}
\end{array}\right)
\left(\begin{array}{c}V_1-V_2\\\bar{V}-V_2\end{array}\right),
\end{equation}
so that the conductance $G_{\rm NS}=\partial
I_1/\partial(V_1-V_2)$ is given by
\begin{equation}\label{G_NS}
G_{\rm NS} = G_{11} - \frac{\left(G_{13}+ G_{14} \right)
\left(G_{31}+ G_{41} \right)}{ G_{33}+G_{34}+G_{43}+G_{44}}.
\end{equation}
Despite the minus signs in Eqs. (\ref{G_N}) and (\ref{G_NS}), both
$G_{\rm N}$ and $G_{\rm NS}$ are positive definite which follows
from the unitarity, time-reversal, and electron hole symmetries of
the S--matrices \cite{datta99}. Eqs. (\ref{G_N}) and (\ref{G_NS})
together with Eqs. (\ref{S_N}), (\ref{See}), and (\ref{She}) form
the basis for our calculations of the conducting properties of
two-probe semiconductor-superconductor junctions with a normal
region containing two scattering regions separated by a
phase-coherent region which couples to a voltage probe.

The sequential tunneling limit ($\zeta=1$) corresponds to series
connected resistors, see panel (d) in Fig. \ref{FIG1}. Thus we can
in general combine Eqs. (\ref{BEENAKKER}) and (\ref{LANDAUER}) to
obtain
\begin{equation}\label{seq}
G_{\rm N} \stackrel{\zeta\rightarrow 1}{=}G_0
\left(\frac{1}{\sum_n T_{1,n}}+\frac{1}{\sum_n
T_{2,n}}\right)^{-1}\,,\,G_{\rm
NS}\stackrel{\zeta\rightarrow 1}{=}G_0 \left(\frac{1}{\sum_n
T_{1,n}} +\frac{1}{2\sum_n
\frac{T_{2,n}^2}{(2-T_{2,n})^2}}\right)^{-1},
\end{equation}
where $T_{n,1}$ and $T_{n,2}$ are the transmission eigenvalues of
$t_1t_1^\dagger$ and $t_2t_2^\dagger$, respectively.

The conductance in the zero-coupling limit ($\zeta=0$) can of
course be calculated directly from Eqs. (\ref{BEENAKKER}) and
(\ref{LANDAUER}) with the transmission eigenvalues $T_{n,12}$ of
$t_{12}t_{12}^\dagger$, where $t_{12}$ is the transmission matrix
of the composite S--matrix ${\cal S}_{12}={\cal S}_2\otimes U
\otimes {\cal S}_1$, see {\it e.g.} the structure of Eq. (\ref{S_1}).

\section{Double-barrier tunneling}

As a specific example we consider double-barrier tunneling with
two identical barriers which we model by delta function potentials
$V({\bf
  r})=H\left[\delta(z)+\delta(z+L)\right]$. Here $L$ is the separation
of the two barriers and $H$ is the strength of the potentials. The
S--matrices are given by
\begin{equation}
{\cal S}_1={\cal S}_2 =\left(\begin{array}{cc}
r_{\scriptscriptstyle\rm \delta} & t_{\scriptscriptstyle\rm
\delta}\\ t_{\scriptscriptstyle\rm
\delta}&r_{\scriptscriptstyle\rm \delta}
\end{array}\right)\,,\,
\left(t_{\scriptscriptstyle\rm\delta}\right)_{nn'} =\delta_{nn'}
\frac{1}{1 + i Z/\cos\theta_n }\,,\,
\left(r_{\scriptscriptstyle\rm\delta}\right)_{nn'} =\delta_{nn'}
\frac{-iZ/\cos\theta_n }{1 + iZ/\cos\theta_n  },
\end{equation}
where $Z\equiv H/\hbar v_{\rm F}$ is a normalized barrier strength
and $\cos\theta_n\equiv k_n/k_{\rm F}$ takes the transverse
momentum of mode $n$ into account
\cite{mortensen99a,mortensen99b}.

\begin{figure}
\begin{center}
\epsfig{file=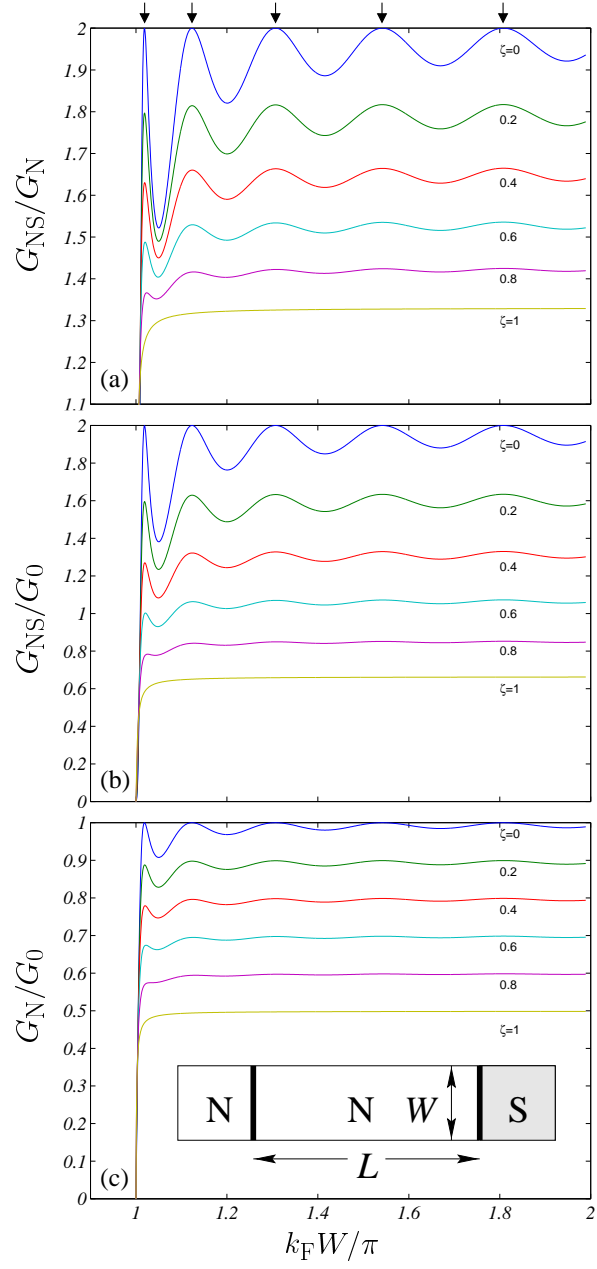, width=0.55\columnwidth}
\end{center}
\caption[]{Double-barrier junction with a normalized barrier
  strength $Z=0.05$ and an aspect ratio $L/W=3$. Panel (c) shows the
  normal state conductance $G_{\rm N}$, (b) the conductance $G_{\rm
    NS}$ of the superconducting state, and (a) the normalized
  conductance $G_{\rm NS}/G_{\rm N}$ for different values of the
  dephasing parameter $\zeta$. The plots are shown as a function of
  $k_{\rm F} W/\pi <2$ corresponding to only a single propagating
  mode. Notice the different scales on the conductance axes.}
\label{FIG2}
\end{figure}

In the phase-coherent limit we calculate the transmission matrix
$t_{12}$ corresponding to the composite S--matrix ${\cal
S}_{12}={\cal
  S}_2 \otimes U\otimes {\cal S}_1$. Since the transmission is
diagonal, $t_{12}t_{12}^\dagger$ has transmission eigenvalues
given by

\begin{equation}
T_{12,n}=\Big(1+2Z_e^2(\theta_n) \big[1+\cos\vartheta_n
+2Z_e(\theta_n)\sin\vartheta_n
  + Z_e^2(\theta_n) \left(1-\cos\vartheta_n\right)
  \big]\Big)^{-1},
\end{equation}
where $Z_e(\theta_n)\equiv Z/\cos\theta_n$ and $\vartheta_n\equiv
2k_{\rm F}L \cos\theta_n$.  Comparing to the transmission
eigenvalues $T_{1,n}=T_{2,n}=\left[1+Z_e^2(\theta_n)\right]^{-1}$
of the individual barriers it is seen that for the double-barrier
system, $T_{12,n}$ can be equal to unity even for a finite value
of $Z$ which is in contrast to the result for $T_{1,n}=T_{2,n}$.
This is one of the important differences between phase-coherent
quantum transmission and classical transmission probabilities.  At
the values of $Z$ where mode $n$ becomes fully transparent a peak
in the conductance is to be expected. Apart from the trivial
solution $Z=0$ we find that $Z=- \cos\theta_n\,
\cot(\vartheta_n/2)$ which means that the mode can be fully
transparent for $\pi \leq \vartheta_n \leq 2\pi$ (modulo $2\pi$)
only.

As a numerical example we consider a double-barrier junction with
an aspect ratio $L/W=3$. Fig. \ref{FIG2} shows the conductance as
a function of $k_{\rm F}W/\pi$ for $Z=0.05$. For $k_{\rm
F}W/\pi<1$ there are no propagating modes and for $1< k_{\rm
F}W/\pi<2$ a single mode is propagating. In panel (c), the normal
state conductance $G_{\rm N}$ increases from zero to approximately
$G_0$ at the onset of the first propagating mode. As the coupling
parameter is increased from $0$ towards $1$, $G_{\rm N}$
approaches $G_0/2$ and it actually becomes a little lower than
$G_0/2$ due to the finite barrier strength (at $k_{\rm F}W/\pi=2$,
Eq. (\ref{seq}) gives $G_{\rm N}/G_0 = 150/301$). In the
coherent regime, oscillations caused by size-resonances are seen
and as the sequential tunneling regime is approached these
resonances vanish.  In panel (b), the conductance $G_{\rm NS}$ in
the superconducting state shows the same overall behavior as seen
for $G_{\rm N}$, even though the size-resonances are much more
pronounced. These size-resonances are even more pronounced in the
normalized conductance $G_{\rm NS}/G_{\rm N}$ shown in panel (a).
As the coupling parameter is increased the resonances vanish and
the normalized conductance is suppressed below the ideal
factor-of-two for the enhancement of $G_{\rm NS}$ compared to
$G_{\rm N}$. For these parameters, ideal transmission and maxima
in the conductance are expected for $k_{\rm F}W/\pi \simeq 1.018,
1.113, 1.296, 1.531, 1.808$ which is in full agreement with the
plots in Fig. \ref{FIG2} where these numbers are indicated by
arrows.

\begin{figure}
\begin{center}
\epsfig{file=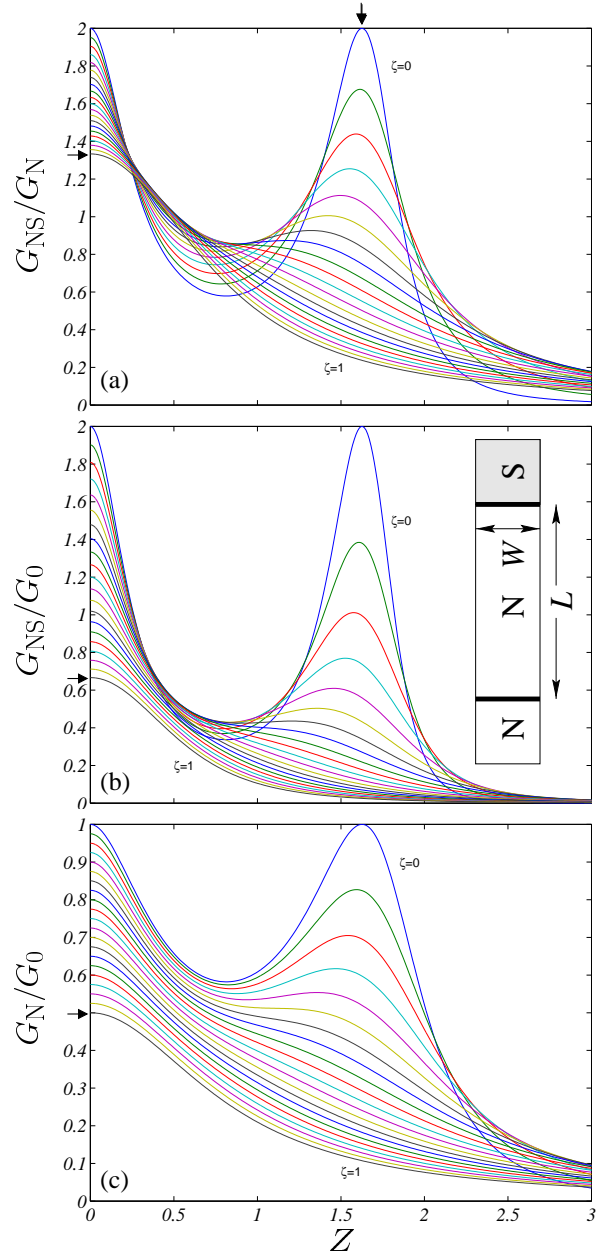, width=0.55\columnwidth}
\end{center}
\caption[]{Double-barrier junction with an aspect ratio $L/W=3$
and a Fermi
  wave vector corresponding to $k_{\rm F} W/\pi=1.9$. Panel (c) shows
  the normal state conductance $G_{\rm N}$, (b) the conductance
  $G_{\rm NS}$ in the superconducting state, and (c) the normalized
  conductance $G_{\rm NS}/G_{\rm N}$ as a function of the normalized
  barrier strength $Z$. The dephasing parameter $\zeta$ is varied from
  $0$ to $1$ in steps of $0.05$. Notice the different scales on the
  conductance axes.}
\label{FIG3}
\end{figure}

In Fig. \ref{FIG3} we consider the $Z$-dependence of the
conductance for a Fermi wave vector corresponding to $k_{\rm F}
W/\pi=1.9$ so that only one mode is propagating. Panels (b) and
(c) show the conductance $G_{\rm NS}$ in the superconducting state
and the normal state conductance $G_{\rm N}$, respectively, as a
function of the normalized barrier strength $Z$ for different
values of the coupling parameter $\zeta$. For these parameters, a
peak in the conductance is expected for $Z\simeq 1.626$ which is
indeed what is clearly seen in the coherent tunneling limit
($\zeta=0$) where this number is indicated by an arrow on the
$Z$-axis. As the sequential tunneling regime is approached
($\zeta\rightarrow 1$) the peak is suppressed and an overall
lowering of the conductance is seen.  However, dephasing may also
suppress a destructive interference in the coherent transmission
resulting in an increased conductance. In this particular case
this is seen for $Z>2.2$ in both $G_{\rm N}$ and $G_{\rm NS}$ and
also in a range around $Z\sim 0.7$ in $G_{\rm NS}$ where the
conductance curves are crossing.  Panel (a) shows the
corresponding normalized conductance $G_{\rm NS}/G_{\rm N}$ which
has the value of two for ideal Andreev scattering and decreases
below the ideal factor-of-two enhancement when $Z$ is increased so
that normal scattering becomes dominant.

For $\zeta=1$ and $Z=0$ Eq. (\ref{seq}) gives that $G_{\rm N}/G_0=1/2$, $G_{\rm NS}/G_0=2/3$, and $G_{\rm
NS}/G_{\rm
  N}=4/3$. These numbers are born out by the numerical calculations as
indicated by arrows on the conductance axes in Fig. \ref{FIG3} and
they can also be seen in Fig. \ref{FIG2} where there is a very low
but finite barrier strength.

In experiments, the barrier strength will often not be a tunable
parameter. However, for a given barrier strength the conductance
can be lowered or enhanced due to phase-coherent quantum
interferences in the double-barrier junction. Thus, increasing the
coupling to the voltage probe, the dephasing can give rise to both
an increasing or decreasing conductance depending on the specific
parameters of the double-barrier junction.

\section{Discussion and conclusion}

In semiconductor structures and also semiconductor-superconductor
junctions the conducting properties depend strongly on whether the
length scale $L$ of the device is much small than the
phase-correlation length $L_c$ or it is comparable to $L_c$.
Another dephasing mechanism could be a coupling to a voltage
probe. In this paper we have studied the effect of coupling a
voltage probe to two-probe semiconductor-superconductor structures
with two scattering potentials in the normal region.

As a particular example we have studied the cross-over from
phase-coherent to sequential tunneling in a double-barrier
semiconductor-superconductor junction. For a finite coupling to
the voltage probe, we find that size-resonances of course become
less pronounced. This may increase or decrease the conductance
when increasing the temperature depending on the specific
parameters of the double-barrier junction. However, we also find
an overall lowering of the conductance and a suppression of the
ideal factor-of-two enhancement of the conductance compared to the
normal state conductance.

\section*{Acknowledgements}

We gratefully acknowledge C.W.J. Beenakker, M. B\"{u}ttiker, and
S. Datta for insightful comments.


\begin{thebibliography}{10}

\bibitem{andreev64}
A.~F. Andreev, Zh. Eksp. Teor. Fiz. {\bf 46}, 1823 (1964) [Sov. Phys. JETP {\bf
  19}, 1228 (1964)].

\bibitem{degennes66}
P.~G. de~Gennes, {\em Superconductivity of Metals and Alloys} (Benjamin, New
  York, 1966).

\bibitem{beenakker92}
C.~W.~J. Beenakker, Phys. Rev. B {\bf 46},  12841  (1992).

\bibitem{landauer57}
R. Landauer, Phil. Mag. {\bf 21},  863  (1970).

\bibitem{landauer70}
R. Landauer, IBM J. Res. Dev. {\bf 1},  223  (1957).

\bibitem{fisher81}
D.~S. Fisher and P.~A. Lee, Phys. Rev. B {\bf 23},  6851  (1981).

\bibitem{buttiker86a}
M. B\"{u}ttiker, Phys. Rev. Lett. {\bf 57},  1761  (1986).

\bibitem{blonder82}
G.~E. Blonder, M. Tinkham, and T.~M. Klapwijk, Phys. Rev. B {\bf 25},  4515
  (1982).

\bibitem{buttiker86b}
M. B\"{u}ttiker, Phys. Rev. B {\bf 33},  3020  (1986).

\bibitem{buttiker88}
M. B\"{u}ttiker, IBM J. Res. Dev. {\bf 32},  63  (1988).

\bibitem{chang97}
L.-F. Chang and P.~F. Bagwell, Phys. Rev. B {\bf 55},  12678  (1997).

\bibitem{gramespacher99}
T. Gramespacher and M. B\"{u}ttiker, Phys. Rev. B {\bf 61},  8125  (2000).

\bibitem{beenakker99}
C.~W.~J. Beenakker, private communication.

\bibitem{lambert93}
C.~J. Lambert, V.~C. Hui, and S.~J. Robinson, J. Phys.: Condens. Matter {\bf
  5},  4187  (1993).

\bibitem{beenakker95}
C.~W.~J. Beenakker,  in {\em Mesoscopic Quantum Physics}, edited by E.
  Akkermans, G. Montambaux, J.-L. Pichard, and J. Zinn-Justin (North--Holland,
  Amsterdam, 1995).

\bibitem{beenakker97}
C.~W.~J. Beenakker, Rev. Mod. Phys. {\bf 69},  731  (1997).

\bibitem{lambert98}
C.~J. Lambert and R. Raimondi, J. Phys.: Condens. Matter {\bf 10},  901
  (1998).

\bibitem{datta95}
S. Datta, {\em Electronic Transport in Mesoscopic Systems} (Cambridge
  University Press, Cambridge, 1995).

\bibitem{takane92}
Y. Takane and H. Ebisawa, J. Phys. Soc. Jpn. {\bf 61},  1685  (1992).

\bibitem{datta99}
Unitarity yields $\sum_{q=1}^M S_{pq}^{\scriptscriptstyle\rm
  ee}\left\{S_{pq}^{\scriptscriptstyle\rm
  ee}\right\}^\dagger+S_{pq}^{\scriptscriptstyle\rm
  eh}\left\{S_{pq}^{\scriptscriptstyle\rm eh}\right\}^\dagger=\hat{1} $
  ($\hat{1}$ being an $N_p\times N_p$ unit matrix) and time-reversal symmetry
  together with the electron-hole symmetry yield ${\rm
  Tr}\,S_{pq}^{\scriptscriptstyle\rm ee}\left\{S_{pq}^{\scriptscriptstyle\rm
  ee}\right\}^\dagger={\rm Tr}\,S_{qp}^{\scriptscriptstyle\rm
  ee}\left\{S_{qp}^{\scriptscriptstyle\rm ee}\right\}^\dagger$ and ${\rm
  Tr}\,S_{pq}^{\scriptscriptstyle\rm eh}\left\{S_{pq}^{\scriptscriptstyle\rm
  eh}\right\}^\dagger={\rm Tr}\,S_{qp}^{\scriptscriptstyle\rm
  eh}\left\{S_{qp}^{\scriptscriptstyle\rm eh}\right\}^\dagger$ (at the Fermi
  level in zero magnetic field); We are grateful to S. Datta (private
  communication) for pointing out the usefulness of these relations in checking
  the numerics; See also Refs.
  \cite{lambert93,beenakker95,beenakker97,lambert98}.

\bibitem{mortensen99a}
N.~A. Mortensen, K. Flensberg, and A.-P. Jauho, Phys. Rev. B {\bf 59},  10176
  (1999).

\bibitem{mortensen99b}
N.~A. Mortensen, K. Flensberg, A.-P. Jauho, and H. Schomerus, Phys. Rev. B {\bf
  60},  13762  (1999).

\end{thebibliography}
\end{document}